# Generalized $h$-index for Disclosing Latent Facts in Citation Networks

Antonis Sidiropoulos[1]    Dimitrios Katsaros[1,2]    Yannis Manolopoulos[1]

[1]Informatics Dept., Aristotle University, Thessaloniki, Greece

[2]Computer & Communications Engineering Dept., University of Thessaly, Volos, Greece

{antonis,dimitris,manolopo}@delab.csd.auth.gr

**Abstract**

What is the value of a scientist and its impact upon the scientific thinking? How can we measure the prestige of a journal or of a conference? The evaluation of the scientific work of a scientist and the estimation of the quality of a journal or conference has long attracted significant interest, due to the benefits from obtaining an unbiased and fair criterion. Although it appears to be simple, defining a quality metric is not an easy task. To overcome the disadvantages of the present metrics used for ranking scientists and journals, J. E. Hirsch proposed a pioneering metric, the now famous *h-index*. In this article, we demonstrate several inefficiencies of this index and develop a pair of generalizations and effective variants of it to deal with scientist ranking and with publication forum ranking. The new citation indices are able to disclose trendsetters in scientific research, as well as researchers that constantly shape their field with their influential work, no matter how old they are. We exhibit the effectiveness and the benefits of the new indices to unfold the full potential of the *h-index*, with extensive experimental results obtained from DBLP, a widely known on-line digital library.

## 1   Introduction

The evaluation of the scientific work of a scientist has long attracted significant interest, due to the benefits from obtaining an unbiased and fair criterion. Having defined such a metric we can use it for faculty recruitment, promotion, prize awarding, funding allocation, comparison of personal scientific merit, etc. Similarly, the estimation of a publication forum's (journal or conference) quality is of particular interest, since it impacts the scientists' decisions about where to publish their work, the researchers' preference in seeking for important articles, and so on.

Although, the issue of ranking a scientist or a journal/conference dates back to the seventies with the seminal work of Eugene Garfield [14] and continued with sparse publications e.g., [15, 17], during the last five years we have witnessed a blossom of this field [3, 5, 6, 18, 19, 21, 22, 23, 24, 27, 29] due to the proliferation of digital libraries, which made available a huge amount of bibliographic data.

Until present there are two major popular ways for evaluating scientific work, and a hybrid of them. The first method is by allowing some contacted experts to perform the ranking and the second method is based on what is termed *citation analysis*, which involves examining an "item"'s (scientist/journal/conference) referring articles. An amalgamation of them is also possible, although it is more close to the latter approach.

The first method adopts an ad hoc approach, which works by collecting the opinion of different experts (or not) in a domain. The study reported in [22] focused in the area of Information Systems and performed an on-line survey for 87 journals with 1000 respondents approximately, whereas the authors of [21] conducted the most extensive survey to date of IS journal rankings. They collected responses from 2559 respondents (32% of the 8741 targeted faculty members in 414 IS departments worldwide). Instead of using a predetermined journal list, they asked the respondents to freely nominate their top-four research journals. This kind of works is very interesting, because they perform a ranking according to readers' (and authors') perception, which is not always adequately expressed through citation analysis, but they suffer from the fact of being basically "manual" sometimes biased, and not highly computerized (automated) and objective.

On the other hand, the second way of evaluating the scientific work is by defining an objective function that calculates some "score" for the "objects" under evaluation, taking into account the graph structure created by the citations among the published articles. Defining a quality and representative metric is not an easy task, since it should account for the productivity of a scientist and the impact of all of his/her work (analogously for journals/conferences). Most of the existing methods up-to-date are based on some form of (arithmetics upon) the total number of authored papers, the average number of authored papers per year, the total number of citations, the average number of citations per paper, the average number of citations per year, etc. A comprehensive description of many of them can be found at [30].

Finally, characteristic works implementing the hybrid approach of combining the experts' judge and citation analysis are described in [19, 32]. Their rankings are realized by taking some averages upon the results obtained from the citation analysis and experts' opinion, thus implementing a post-processing step of the two major approaches.

## 1.1 Motivation for new citation indices

Although, there is no clear winner among citation analysis and experts' assessment, the former method is usually the preferred method, because it can be performed in a fully automated and computerized manner and it is able to exploit the wealth of citation information available in digital libraries.

All the metrics used so far in citation analysis, even those which are based on popular spectral techniques, like HITS [20], PageRank [25] and its variations for bibliometrics, like [7], present one or more of the following drawbacks (see also [16]):

- They do not measure the importance or impact of papers, e.g., the metrics based solely on the total number of papers.

- They are affected by a small number of "big hits" articles, which received huge number of citations, whereas the rest of the articles may have negligible total impact, e.g., the metrics based on the total number of citations.

- They can not measure productivity, e.g., the metrics based on the average number of citations per paper.

- They have difficulty to set administrative parameters, e.g., the metrics based on the number $x$ of articles, which have received $y$ citations each, or the metrics based on the number $z$ of the most cited articles.

To collectively overcome all these disadvantages of the present metrics, last year J. E. Hirsch proposed the pioneering *h-index* [16], which, in a short period of time, became extremely popular[1]. The *h-index* is defined as follows:

**Definition 1.** *A researcher has h-index h if h of his/her $N_p$ articles have received at least h citations each, and the rest $(N_p - h)$ articles have received no more than h citations [1, 16].*

This metric calculates how broad the research work of a scientist is. The *h-index* accounts for both productivity and impact. For some researcher, to have large *h-index*, s/he must have a lot of "good" articles, and not just a few "good" articles.

The *h-index* acts as a lower bound on the real number of citations for a scientist. In fact, there is a significant gap between the total number of citations as calculated by *h-index* and the real number of citations of a scientist. Think that the quantity $h$ will always be smaller than or equal to the number $N_p$ of the articles of a researcher; it holds that $h^2 \leq N_{c,tot}$, where $N_{c,tot}$ is the total number of citations that the researcher has received. Apparently, the equality holds when all the articles, which contribute to *h-index* have received exactly $h$ citations each, which is quite improbable. Therefore, in the usual case it will hold that $h^2 < N_{c,tot}$. To bridge this gap, J. E. Hirsch defined the index $a$ as follows:

**Definition 2.** *A scientist has a-index a if the following equation holds [16]:*

$$N_{c,tot} = ah^2. \tag{1}$$

---

[1]Notice that the economics literature defines the $H$-index (the Herfindahl-Hirschman index), which is a way of measuring the concentration of market share held by particular suppliers in a market. The $H$ index is the sum of squares of the percentages of the market shares held by the firms in a market. If there is a monopoly, i.e., one firm with all sales, the $H$ index is 10000. If there is perfect competition, with an infinite number of firms with near-zero market share each, the $H$ index is approximately zero. Other industry structures will have $H$ indices between zero and 10000.

The *a-index* can be used as a second metric-index for the evaluation and ranking of scientists. It describes the "magnitude" of each scientist's "hits". A large $a$ implies that some article(s) have received a fairly large number of citations compared to the rest of its articles and with respect to what the *h-index* presents.

The introduction of the *h-index* was a major breakthrough in citation analysis. Though several aspects of the inefficiency of the original *h-index* are apparent; or to state it in its real dimension, significant efforts are needed to unfold the full potential of *h-index*. Firstly, the original *h-index* assigns the same importance to all citations, no matter what their age is, thus refraining from revealing the trendsetters scientists. Secondly, the *h-index* assigns the same importance to all articles, thus making the young researchers to have a relatively small *h-index*, because they did not have enough time either to publish a lot of good articles, or time to accumulate large number of citation for their good papers. Thus, the *h-index* can not reveal the brilliant though young scientists.

## 1.2 Our contributions

The purpose of our work is to extend and generalize the original *h-index* in such ways, so as to reveal various latent though strong facts hidden in citation networks. Our proposals aim to maintain the elegance and ease of computation of the original *h-index*, thus we strive for developing relatively simple indexes, since we believe that the simplicity of the *h-index* is one of its beauties. In this context, the article makes the following contributions:

- Introduces two generalizations of the *h-index*, namely the *contemporary h-index* and the *trend h-index*, which are appropriate for scientist ranking and are able to reveal *brilliant young scientists* and *trendsetters*, respectively. These two generalizations can also be used for the cases of conference and journal ranking.

- Introduces a normalized version of the *h-index* for scientist ranking, namely the *normalized h-index*.

- Introduces two variants of the *h-index* appropriate for journal/conference ranking, namely the *yearly h-index* and the *normalized yearly h-index*.

- Performs an extensive experimental evaluation of the aforementioned citation indices, using real data from DBLP, an online bibliographic database.

Developing mathematical models and conducting theoretical analysis of the properties of the proposed indexes is the next step in this work, but it is beyond the scope of this paper; here we are interesting in providing extensive experimental evidence of the power of the generalizations to the *h-index*.

The rest of this article is organized as follows: In Section 2, we present the novel citation indices which are devised for scientist ranking. Section 3 presents the citation indices extending the *h-index* for journal/conference ranking. We present the evaluation of the introduced citation indices in Section 4 and finally, Section 5 summarizes the paper contributions and concludes the article.

## 2 Novel Citation Indices for Scientist Ranking

After the introduction of the *h-index*, a number of other proposals followed, either presenting case studies using it [2, 4, 9, 10, 26, 28], or describing a new variation of it [12] (aiming to bridge the gap between the lower bound of total number of citations calculated by *h-index* and their real number), or studying its mathematics and its performance [8, 11].

Deviating from their line of research, we develop in this article a pair of generalizations of the *h-index* for ranking scientists, which are novel citation indices, a normalized variant of the *h-index* and a pair of variants of the *h-index* suitable for journal/conference ranking.

**The contemporary *h-index*.**
The original *h-index* does not take into account the "age" of an article. It may be the case that some scientist contributed a number of significant articles that produced a large *h-index*, but now s/he is rather inactive or retired. Therefore, senior scientists, who keep contributing nowadays, or brilliant young scientists, who are expected to contribute a large number of significant works in the near future but now they have only a small number of important articles due to the time constraint, are not distinguished by the original *h-index*. Thus, arises the need to define a generalization of the *h-index*, in order to account for these facts.

We define a novel score $S^c(i)$ for an article $i$ based on citation counting, as follows:

$$S^c(i) = \gamma * (Y(now) - Y(i) + 1)^{-\delta} * |C(i)| \qquad (2)$$

where $Y(i)$ is the publication year of article $i$ and $C(i)$ are the articles citing the article $i$. If we set $\delta=1$, then $S^c(i)$ is the number of citations that the article $i$ has received, divided by the "age" of the article. Since, we divide the number of Citations with the time interval, the quantities $S^c(i)$ will be too small to create a meaningful *h-index*; thus, we use the coefficient $\gamma$. In our experiments, reported in Sections 4, we use the value of 4 for the coefficient $\gamma$. Thus, for an article published during the current year, its citations account four times. For an article published 4 year ago, its citations account only one time. For an article published 6 year ago, its citations account $\frac{4}{6}$ times, and so on.

This way, an old article gradually loses its "value", even if it still gets citations. In other words, in the calculations we mainly take into account the newer articles[2]. Therefore, we define a novel citation index for scientist rankings, the *contemporary h-index*, expressed as follows:

**Definition 3.** *A researcher has contemporary h-index $h^c$, if $h^c$ of its $N_p$ articles get a score of $S^c(i) \geq h^c$ each, and the rest $(N_p - h^c)$ articles get a score of $S^c(i) \leq h^c$.*

### The trend *h-index*.

The original *h-index* does not take into account the year when an article acquired a particular citation, i.e., the "age" of each citation. For instance, consider a researcher who contributed to the research community a number of really brilliant articles during the decade of 1960, which, say, got a lot of citations. This researcher will have a large *h-index* due to the works done in the past. If these articles are not cited anymore, it is an indication of an outdated topic or an outdated solution. On the other hand, if these articles continue to be cited, then we have the case of an *influential mind*, whose contributions continue to shape newer scientists' minds. There is also a second very important aspect in aging the citations. There is the potential of disclosing *trendsetter*, i.e., scientists whose work is considered pioneering and sets out a new line of research that currently is hot ("trendy"), thus this scientist's works are cited very frequently.

To handle this, we take the opposite approach than *contemporary h-index*'s; instead of assigning to each scientist's article a decaying weight depending on its age, we assign to each citation of an article an exponentially decaying weight, which is as a function of the "age" of the citation. This way, we aim at estimating the impact of a researcher's work in a particular time instance. We are not interested in how old the articles of a researcher are, but whether they still get citations. We define an equation similar to Equation 2, which is expressed as follows:

$$S^t(i) = \gamma * \sum_{\forall x \in C(i)} (Y(now) - Y(x) + 1)^{-\delta} \qquad (3)$$

where $\gamma$, $\delta$, $Y(i)$ and $S(i)$ for an article $i$ are as defined earlier. We define a novel citation index for scientist ranking, the *trend h-index*, expressed as follows:

**Definition 4.** *A researcher has trend h-index $h^t$ if $h^t$ of its $N_p$ articles get a score of $S^t(i) \geq h^t$ each, and the rest $(N_p - h^t)$ articles get a score of $S^t(i) \leq h^t$ each.*

Apparently, for $\gamma = \delta = 1$, the *trend h-index* coincides with the original *h-index*.

It is straightforward to devise a generalization of both the *contemporary h-index* and *trend h-index*, which takes into account both the age of a scientist's article and the age of each citation to this article, but such index does not provide many additional insights about the real contributions of a scientist. Therefore, we dot not investigate further this generalization in the present article.

### The normalized *h-index*.

Since the scientists do not publish the same number of articles, the original *h-index* is not the fairer metric; thus, we define a normalized version of *h-index*, expressed as follows:

---

[2] Apparently, if $\delta$ is close to zero, then the impact of the time penalty is reduced, and, for $\delta = 0$, this variant coincides with the original *h-index* for $\gamma = 1$.

**Definition 5.** *A researcher has normalized h-index $h^n = h/N_p$, if h of its $N_p$ articles have received at least h citations each, and the rest $(N_p - h)$ articles received no more than h citations.*

In the next section, we define some variants of the *h-index* family of citation indices for ranking journals/conferences.

## 3 New Citation Indices for Journals and Conferences Ranking

Based on the original idea of the *h-index* and on the aforementioned generalizations and variants, we define analogous concepts for ranking journals and conferences. For instance, the *h-index* of a journal/magazine or of a conference is $h$, if $h$ of the $N_p$ articles it contains, have received at least $h$ citations each, and the rest $(N_p - h)$ articles received no more than $h$. The generalizations of *contemporary h-index* and *trend h-index* can be defined for conferences and journals as well similarly to the Definitions 3 and 4. Direct applications of the *h-index* in journal ranking following this definition appeared in [2, 9, 28]. Though, we observe that the direct application of the index can not guarantee a fair comparison between conferences or between journals, because a) their lives are different, and b) they publish different numbers of articles.

We deal with the first problem by calculating the *h-index* on a per year basis. In particular, we define that:

**Definition 6.** *A conference or a journal has yearly h-index $h_y$ for the year y if $h_y$ of its articles $N_{p,y}$ published during the year y have received at least $h_y$ citations each, and the rest $(N_{p,y} - h_y)$ articles received no more than $h_y$ citations.*

For instance, the $h$ index for the year 1992, denoted as $h_{1992}$, of the conference *VLDB* is computed as the number of its articles which have received more than $h_{1992}$ citations. The drawbacks though of the aforementioned metric are the following:

1. The conferences/journals do not publish exactly the same number of articles. Thus, for a conference which published around 50 articles, the upper bound for its *h*-index is 50. Another conference which published 150 the upper bound for its *h*-index is 150, and it also has much more stronger probability to exceed the limit of 50. The number of articles appearing in a year in a conference or journal reflects the preference of the researchers to this publication forum. If we consider that the forum published 50 articles, because it could not attract more valuable articles, then it correctly has as upper bound the number 50 and it is not a problem that it can not overrule forum $B$. On the other hand, perhaps we are interested in the average "quality" of the articles published in a forum, no matter what the number of published articles in a forum is.

2. The $h_y$ index constantly changes. Even though we examine a conference which took place in 1970, the $h_y$ index that we can calculate today, is possible to change a few year later. Thus, the drawback of this index is that we can not have a final evaluation for the forums of a year, no matter how old are they.

The only way to overcome the second drawback, is to add a time window after the organization of a conference or the publication of a journal (i.e., ten or five years time window). This would add the notion of the Impact Factor [31] to the metric, which is beyond the scope of our current research.

To address the first drawback, we define a "parallel" index, which is normalized with respect to the number of articles published in a forum. Its formal definition is given below:

**Definition 7.** *A conference or journal for the year y has **normalized index** $h_y^n = h_y/N_{p,y}$, if $h_y$ of its $N_{p,y}$ articles in the year y have received at least $h_y$ citations each, and the rest $(N_{p,y} - h_y)$ articles received no more than $h_y$ citations.*

Having defined these generalizations and variants of the original *h-index*, we will evaluate in the subsequent sections their success in identifying scientists or forums with extraordinary performance or their ability to reveal latent facts in a citation network, such as brilliant young scientists and trendsetters. For the evaluation, we will exploit the on-line database of DBLP [3].

---
[3]The DBLP digital library with bibliographic data on "Databases and Logic Programming" is maintained by Michael Ley at the University of Trier, accessible from http://dblp.uni-trier.de/

# 4   Experiments

In this section we will present the ranking results for scientists, conferences and journals by using the basic *h-index* definition as well as by using the generalizations and variants developed in the previous sections. Along the lines of [30, 31, 32], our dataset consists of the DBLP collection (DBLP timestamp: Mar/3/2006). The reasons for selecting this source of data instead of ISI or Google Scholar or CiteSeer are the following:

1. DBLP contains data about journal and conference publications as well.

2. DBLP data are focused mostly in the area of Databases.

3. The maintainers of DBLP library put a lot of work into resolving the "names problem" - the same person referenced with (many) different names.

4. DBLP's data format gave us the possibility to eliminate "self-citations", which is not done by Google Schoolar.

5. Google Scholar only takes into account papers it finds on the Web.

6. ISI's coverage for computer science is not comprehensive.

7. CiteSeer does not eliminate "self-citations" and does not rank author queries by citation number and also weights them by year.

It is worthwhile noticing that many top conferences of this area are very competitive (with an acceptance ratio stronger than 1:3 and up to 1:7), and occasionally more competitive that the top journals of the area. In many computer science departments worldwide, publications in these conferences are favored in comparison to journal publications. Therefore, a ranking of conferences on databases is equally important to the ranking of the journals of the area.

The used database snapshot contains 451694 inproceedings, 266307 articles, 456511 authors, 2024 conference series and 504 journals. Also, the number of citations in our dataset is 100205. Although this number is relatively small, it is a satisfactory sample for our purposes. Almost all citations in the database are made from publications prior to the year 2001. Thus, we can assume that the results presented here correspond to the year 2001. From now on, with the term "now" we actually mean sometime near 2001. Although other bibliographic sources (e.g., ISI, Google Scholar, CiteSeer) are widely available, the used collection has the aforementioned desired characteristics and thus it is sufficient for exhibiting the benefits of our proposed citation indices, without biasing our results.

## 4.1   Experiments with the h-index for scientists

In Tables 1, 2, 3 and 4 we present the resulting ranking using the *h-index*, as well as its defined generalizations. In these tables columns $a_c$ and $a_t$ stand for the *a-index* corresponding to *contemporary h-index* and *trend h-index*, respectively. The computation of $a_c$ (and $a_t$) is analogous to the original *a-index* computation, but it uses the functions $S_c(i)$ (and $S_t(i)$) defined earlier for all publications of each author, rather than using the total number of citations[4]. This is depicted in Equations 4 and 5 where $P$ is the set of author's publications.

$$\sum_{\forall i \in P} S_c(i) = a_c h_c^2 \qquad (4)$$

$$\sum_{\forall i \in P} S_t(i) = a_t h_t^2 \qquad (5)$$

Based on the *contemporary h-index* and *trend h-index* definitions, it is obvious that the conditions $\sum_{\forall i \in P} S_c(i) \geq h_c^2$ and $\sum_{\forall i \in P} S_t(i) \geq h_t^2$ are true. The equality holds only in the unusual case that every $S_c(i)$ is equal to $h_c$.

At a first glance, we see that the values computed for *h-index* (Table 1) are much lower than the values presented in [16] for physics scientists due to the non completeness of the source data. Also, we can notice that the values for $h, h_c$ and $h_t$ are different from each other as well as there are differences in the ordering of the scientists. This confirms our allegation for the convenience of these indices and will be discussed in the sequel of the article.

---

[4]Notice here that the equations $N_{c,tot} = a_c h_c^2$ and $N_{c,tot} = a_t h_t^2$ are not true since $\sum_{\forall i \in P} S_c(i)$ and $\sum_{\forall i \in P} S_t(i)$ are different than $N_{c,tot}$.

A superficial examination of Tables 2 and 3, does not reveal any major difference between their ranking and the ranking obtained by *h-index* (in Table 1). With respect to Table 2, the astute reader though, will catch three important representative cases: the case of Christos Faloutsos, the case of Serge Abiteboul and the case of Jennifer Widom. Christos Faloutsos is at the $16^{th}$ place of *h-index* table. In *contemporary h-index* table he climbs to the $14^{th}$ position. Serge Abiteboul climbs up from the $13^{th}$ position to the $5^{th}$ position. Similarly, Jennifer Widom appears in the $6^{th}$ position of the *contemporary h-index* (Table 2), although she does not have an entry in the *h-index* (see Table 1). This means that the major amount of their good publications is published in the resent years (relatively to the rest of the scientists). In other words, they work on now hot topics. Consequently, we would characterize their works as *contemporary*.

The results appear more impressive in the *trend h-index* (Table 3). Christos Faloutsos climbs to the $8^{th}$ position, and Jennifer Widom in the $5^{th}$ position. This shows that their publications get citations during the very recent years. Consequently, we would characterize the work of Faloutsos and Widom as *"trendy"*, in the sense that a general interest exists by the rest of the research community for the work of the specific scientists during the particular time period. Indeed, Faloutsos is recognized as (one of) the main trendsetter in the area of spatial, multidimensional and time series data management. Widom is recognized as (one of) the main trendsetter in the area of semistructured data management. Figures 1(a) and 1(e), which will be examined in the sequel, reveal that the main volume of the total citations to these scientists started at specific years, when they contributed significant ground-setting papers in their domains.

| Name | $h$ | $a$ | $N_{c,tot}$ | $N_p$ |
|---|---|---|---|---|
| 1.Michael Stonebraker | 24 | 3.78 | 2180 | 193 |
| 2.Jeffrey D. Ullman | 23 | 3.37 | 1783 | 227 |
| 3.David J. DeWitt | 22 | 3.91 | 1896 | 150 |
| 4.Philip A. Bernstein | 20 | 3.39 | 1359 | 124 |
| 5.Won Kim | 19 | 2.96 | 1071 | 143 |
| 6.Catriel Beeri | 18 | 3.16 | 1024 | 93 |
| 7.Rakesh Agrawal | 18 | 3.06 | 994 | 154 |
| 8.Umeshwar Dayal | 18 | 2.81 | 913 | 130 |
| 9.Hector Garcia-Molina | 17 | 3.60 | 1041 | 314 |
| 10.Yehoshua Sagiv | 17 | 3.52 | 1020 | 121 |
| 11.Ronald Fagin | 17 | 2.83 | 818 | 121 |
| 12.Jim Gray | 16 | 6.13 | 1571 | 118 |
| 13.Serge Abiteboul | 16 | 4.33 | 1111 | 172 |
| 14.Michael J. Carey | 16 | 4.25 | 1090 | 151 |
| 15.Nathan Goodman | 16 | 3.37 | 865 | 68 |
| 16.Christos Faloutsos | 16 | 2.89 | 742 | 175 |
| 17.Raymond A. Lorie | 15 | 6.23 | 1403 | 35 |
| 18.Jeffrey F. Naughton | 15 | 2.90 | 653 | 123 |
| 19.Bruce G. Lindsay | 15 | 2.76 | 623 | 60 |
| 20.David Maier | 14 | 5.56 | 1090 | 158 |

Table 1: Scientist ranking using the *h-index*.

| Name | $h_c$ | $a_c$ | $h$ | $N_{c,tot}$ | $N_p$ |
|---|---|---|---|---|---|
| 1.David J. DeWitt | 14 | 3.10 | 22 | 1896 | 150 |
| 2.Jeffrey D. Ullman | 13 | 3.44 | 23 | 1783 | 227 |
| 3.Michael Stonebraker | 12 | 3.98 | 24 | 2180 | 193 |
| 4.Rakesh Agrawal | 12 | 3.24 | 18 | 994 | 154 |
| 5.Serge Abiteboul | 11 | 4.08 | 16 | 1111 | 172 |
| 6.Jennifer Widom | 11 | 3.23 | 14 | 709 | 136 |
| 7.Jim Gray | 10 | 3.93 | 16 | 1571 | 118 |
| 8.Michael J. Carey | 10 | 3.79 | 16 | 1090 | 151 |
| 9.Won Kim | 10 | 3.00 | 19 | 1071 | 143 |
| 10.David Maier | 10 | 2.93 | 14 | 1090 | 158 |
| 11.Hector Garcia-Molina | 9 | 5.30 | 17 | 1041 | 314 |
| 12.Jeffrey F. Naughton | 9 | 3.85 | 15 | 653 | 123 |
| 13.Yehoshua Sagiv | 9 | 3.76 | 17 | 1020 | 121 |
| 14.Christos Faloutsos | 9 | 3.68 | 16 | 742 | 175 |
| 15.Catriel Beeri | 9 | 3.59 | 18 | 1024 | 93 |
| 16.Philip A. Bernstein | 9 | 3.49 | 20 | 1359 | 124 |
| 17.Umeshwar Dayal | 9 | 3.39 | 18 | 913 | 130 |
| 18.Hamid Pirahesh | 9 | 3.34 | 14 | 622 | 67 |
| 19.H. V. Jagadish | 9 | 2.88 | 12 | 503 | 151 |
| 20.Raghu Ramakrishnan | 8 | 5.05 | 14 | 818 | 147 |

Table 2: Scientist ranking using the *contemporary h-index*.

| Name | $h_t$ | $a_t$ | $h$ | $N_{c,tot}$ | $N_p$ |
|---|---|---|---|---|---|
| 1.David J. DeWitt | 20 | 2.73 | 22 | 1896 | 150 |
| 2.Michael Stonebraker | 17 | 3.61 | 24 | 2180 | 193 |
| 3.Jeffrey D. Ullman | 17 | 3.45 | 23 | 1783 | 227 |
| 4.Rakesh Agrawal | 17 | 3.06 | 18 | 994 | 154 |
| 5.Jennifer Widom | 16 | 2.81 | 14 | 709 | 136 |
| 6.Serge Abiteboul | 14 | 4.07 | 16 | 1111 | 172 |
| 7.Hector Garcia-Molina | 14 | 4.03 | 17 | 1041 | 314 |
| 8.Christos Faloutsos | 14 | 3.15 | 16 | 742 | 175 |
| 9.Jim Gray | 13 | 4.46 | 16 | 1571 | 118 |
| 10.Jeffrey F. Naughton | 13 | 3.36 | 15 | 653 | 123 |
| 11.Won Kim | 13 | 3.23 | 19 | 1071 | 143 |
| 12.Michael J. Carey | 12 | 4.79 | 16 | 1090 | 151 |
| 13.Yehoshua Sagiv | 12 | 3.96 | 17 | 1020 | 121 |
| 14.Umeshwar Dayal | 12 | 3.41 | 18 | 913 | 130 |
| 15.Catriel Beeri | 12 | 3.12 | 18 | 1024 | 93 |
| 16.Raghu Ramakrishnan | 11 | 4.41 | 14 | 818 | 147 |
| 17.Philip A. Bernstein | 11 | 4.03 | 20 | 1359 | 124 |
| 18.David Maier | 11 | 3.94 | 14 | 1090 | 158 |
| 19.Hamid Pirahesh | 11 | 3.87 | 14 | 622 | 67 |
| 20.H. V. Jagadish | 11 | 3.58 | 12 | 503 | 151 |

Table 3: Scientist ranking using the *trend h-index*.

| Name | $h_n$ | $h$ | $a$ | $N_{c_tot}$ | $N_p$ |
|---|---|---|---|---|---|
| 1.Rajiv Jauhari | 1 | 5 | 3.72 | 93 | 5 |
| 2.Jie-Bing Yu | 1 | 5 | 2.36 | 59 | 5 |
| 3.L. Edwin McKenzie | 1 | 5 | 2.04 | 51 | 5 |
| 4.Upen S. Chakravarthy | 0.88 | 8 | 2.60 | 167 | 9 |
| 5.James B. Rothnie Jr. | 0.85 | 6 | 6.55 | 236 | 7 |
| 6.M. Muralikrishna | 0.85 | 6 | 5.47 | 197 | 7 |
| 7.Stephen Fox | 0.83 | 5 | 4.12 | 103 | 6 |
| 8.Antonin Guttman | 0.8 | 4 | 20.43 | 327 | 5 |
| 9.Marc G. Smith | 0.8 | 4 | 4.81 | 77 | 5 |
| 10.Gail M. Shaw | 0.8 | 4 | 4.37 | 70 | 5 |
| 11.Glenn R. Thompson | 0.8 | 4 | 4.37 | 70 | 5 |
| 12.David W. Shipman | 0.75 | 6 | 11.16 | 402 | 8 |
| 13.Dennis R. McCarthy | 0.75 | 6 | 5.30 | 191 | 8 |
| 14.Spyros Potamianos | 0.66 | 4 | 10.43 | 167 | 6 |
| 15.Robert K. Abbott | 0.66 | 4 | 4.68 | 75 | 6 |
| 16.Edward B. Altman | 0.66 | 4 | 3.06 | 49 | 6 |
| 17.Brian M. Oki | 0.66 | 4 | 2.56 | 41 | 6 |
| 18.Gene T. J. Wuu | 0.66 | 6 | 2.25 | 81 | 9 |
| 19.Marguerite C. Murphy | 0.66 | 4 | 1.62 | 26 | 6 |
| 20.Gerald Held | 0.62 | 5 | 9.84 | 246 | 8 |

Table 4: Scientist ranking using the *normalized h-index*.

It is also worthwhile to mention that the *contemporary h-index* and *trend h-index* are fair metrics for the "all-time classic" scientists, e.g., Jeffrey Ullman, Michael Stonebraker, and David DeWitt, whose influential works continue to shape the modern scientists way of thinking.

By examining the Tables 2, 3 and 4, we observe than the rankings provided by *normalized h-index* differ significantly from the rankings provided by the other metrics. This happens because the scientists with few but good publications take advantage. Thus, we cannot evaluate the research work of a scientist by taking into consideration only the *normalized h-index*. The *normalized h-index* can be used in parallel to *h-index* and as a second criterion. We can easily assume that the majority of the scientists presented in Table 4 are probably PhD or MSc students that wrote a fine article with a "famous" professor and after that, they stopped their research career. The second possibility is that the main part of their articles are not included in the DBLP collection – probably because they actually belong to a scientific discipline other than Databases and Logic Programming. Finally, it is always possible to track "promising" researchers among them, who will continue their significant research work.

Motivated by the differences in the above tables, we present the collection of graphs in Figure 1. In these figures, we can see the history of the *h-index* for those scientists, who present significant differences between the *h-index* family of citation indices, and also those who have a rapid upwards slope at their plot curves[5]. For each scientist, we provide a justification for the resulting curves related to his/her main research interests. It is the case that these scientists have really broad interests that can not be confined in a single term. Though, we attempt to characterize the field where they made their most significant contributions.

Comparing Figures 1(a) and 1(b) regarding Christos Faloutsos and Jim Gray respectively, we can see that the two scientists have the same *h-index* now. However, Christos Faloutsos has a more ascending slope than Jim Gray, since he started being cited on 1984, while Jim Gray on 1976. Also, the *trend h-index* ($h_t$) curve of Christos Faloutsos stays constantly over the *h-index* ($h$) equivalent. This means that Christos Faloutsos is getting cited very often at the present and thus, we expect his *h-index* to get higher than Jim Gray's *h-index*. Finally, Jim Gray's *contemporary h-index* ($h_c$) is constantly below $h$ since 1985 and it diverges with time. This indicates that since 1985 he has not presented exceptionally seminal papers (relatively to his older ones) and after this point the progress is digressive. Indeed, the Turing Award recipient Jim Gray made ground setting contributions to transaction processing; this topic is considered as a solved problem now and it is not considered as an active research area. On the other hand, the main contributions of Christos Faloutsos focus in the area of spatial and multidimensional data management, which became very popular roughly since 1995.

Figures 1(c) and 1(d) correspond to Michael Stonebraker and David J. DeWitt. Both of these researchers are on the top of our list. We can notice that David J. DeWitt' *contemporary h-index* is very close to his *h-index*, which means that he keeps publishing very good papers. On the contrary, Michael Stonebraker has started to deflect since 1985. This helps us understand that Michael Stonebraker' *trend h-index* will also decrease after some years, as it is shown in the same figure. Thus, while Michael Stonebraker is in higher position than David J. DeWitt at the *h-index* ranking, David J. DeWitt comes first when examining the other two variations. This means that, if the productivity level of the two researchers keeps on the same pace, the second will soon surpass the first one at the *h-index* as well.

In Figure 1(e), we see the progress rate for Jennifer Widom. While Jennifer Widom is not even among the top 20 researchers using the *h-index*, she is on the $6^{th}$ and $5^{th}$ position using the *contemporary h-index* and *trend h-index*, respectively. She is the only researcher from our list that presents such a big difference on the timing rates compared to the basic *h-index*. As we can also see from the diagram, this difference is justifiable, since the increase rate of the basic *h-index* is high. She is also the only researcher that her *contemporary h-index* is constantly close to *h-index* and not below. Finally, although, for all the researchers that we present, the *trend h-index* is always lower than the *h-index* in the year of 2000, the *trend h-index* stays higher in her case. Jennifer Widom made some ground breaking contributions on building semistructured data management systems, that laid the foundations for the modern XML management systems.

In Figure 1(f), Won Kim presents an analogous path with Stonebraker. For instance, there is a high ascending curve for *trend h-index*, but *contemporary h-index* remains low after 1990, and it is finally obvious that *trend h-index* will also follow a decreasing path. Therefore, we expect that *h-index* will not present high increase. This is explained by the fact that the main research interests of Won Kim was on object-oriented database systems, which flourished during the last years of the eighties and in the first years of the nineties, but later become a relatively inactive area.

Our observations about Figure 1(g) concerning the work of Rakesh Aggrawal are analogous to those on Figure 1(a) concerning Christos Faloutsos, thus we do not go into further details. In Figure 1(h) concerning Yannis Ioannidis, we see an increasing trend that is analogous to that of Jim Gray. The *trend h-index* remains constantly over *h-index*, which means that there is a remarkable potential. In addition, the *contemporary h-index* presents a small deflection from *h-index* after 1993, which is completely analogous to that of Jim Gray after 1985. Based on the available data, Yannis Ioannidis follows the same progress path as Jim Gray, with a time lag of about 10 years.

---

[5]Again, we remind that our data set is rather incomplete for the years after 2000, and thus a downwards pitch for all the researchers appears during the years 1999-2000. However, the results are indicative.

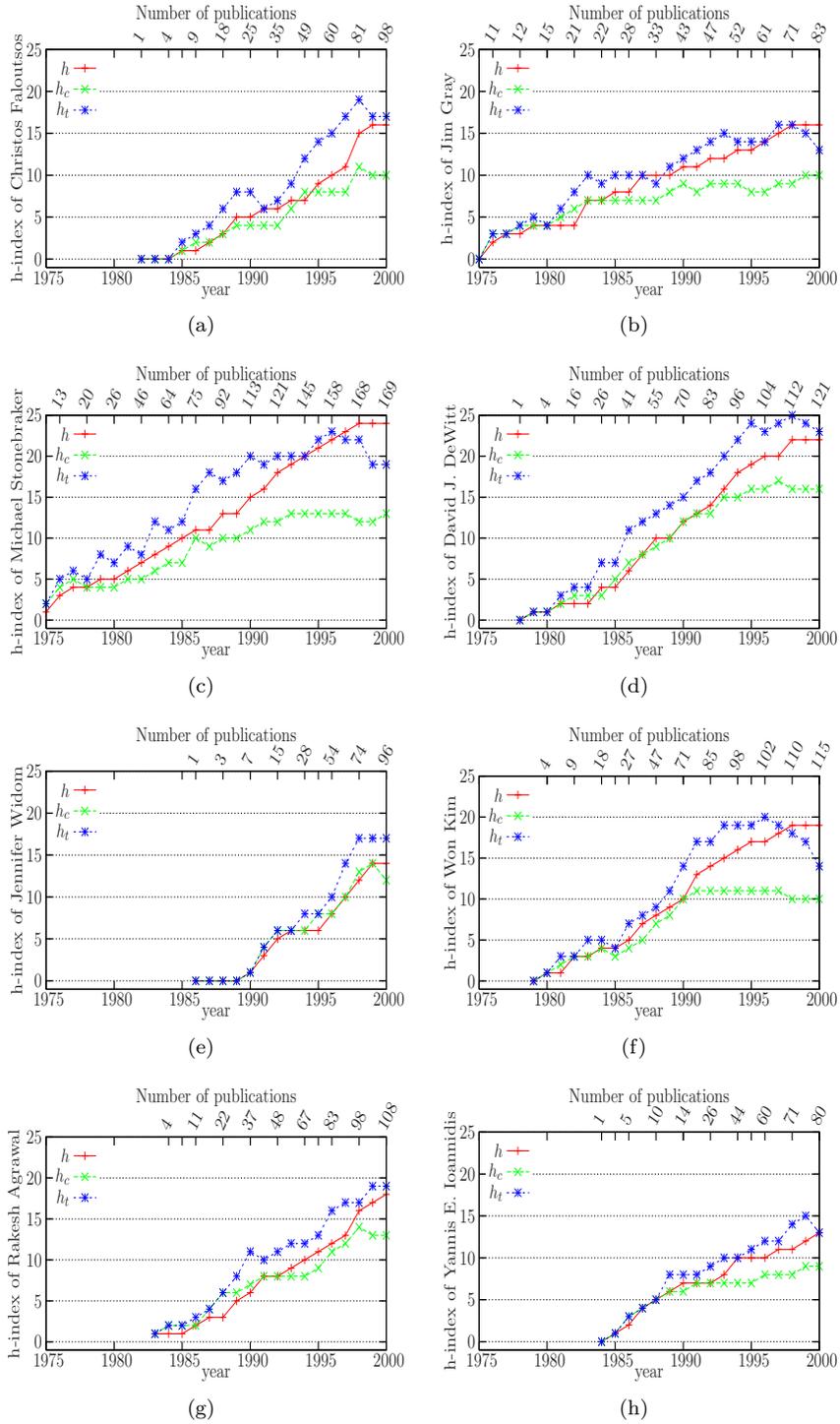

Figure 1: The *h-index* of scientists working in databases area.

### 4.1.1 Matching the *h-index* rankings to the awarded scientists

Our next investigation concerns an experiment to test whether the rankings by the *h-index* and its generalizations are in accordance with the awards for the database scientific domain. In papers [30] and [32] we used the '*SIGMOD E. F. Codd Innovations Award*' to evaluate some ranking methods. Here, we perform an analogous experiment. The higher an awarded scientist is ranked, the better the ranking method is considered to be. We have to note here, that the issue of awarding a scientists a particular award is not simply a matter of arithmetics (some numerical metrics), but it is much complex. Though, we use this methodology to test whether the citation indices defined in this metric reveal some general trends.

| Name | $h$ | $h_c$ | $h_t$ | $h$ | $h_c$ | $h_t$ | year | $h$ | $h_c$ | $h_t$ |
|---|---|---|---|---|---|---|---|---|---|---|
| Michael Stonebraker | 1 | 3 | 2 | 3 | 2 | 1 | 1992 | 1 | 2 | 1 |
| Jim Gray | 12 | 7 | 9 | 12 | 11 | 10 | 1993 | 15 | 11 | 8 |
| Philip A. Bernstein | 4 | 16 | 17 | 2 | 6 | 4 | 1994 | 2 | 9 | 5 |
| David J. DeWitt | 3 | 1 | 1 | 3 | 1 | 1 | 1995 | 2 | 1 | 1 |
| C. Mohan | 28 | 37 | 31 | 44 | 36 | 35 | 1996 | 49 | 23 | 19 |
| David Maier | 20 | 10 | 18 | 11 | 9 | 15 | 1997 | 15 | 10 | 17 |
| Serge Abiteboul | 13 | 5 | 6 | 17 | 4 | 11 | 1998 | 16 | 6 | 11 |
| Hector Garcia-Molina | 9 | 11 | 7 | 10 | 8 | 4 | 1999 | 14 | 7 | 5 |
| Rakesh Agrawal | 7 | 4 | 4 | 9 | 4 | 4 | 2000 | 7 | 4 | 4 |
| Rudolf Bayer | 145 | 196 | 183 | 142 | 218 | 222 | 2001 | 145 | 196 | 183 |
| Patricia G. Selinger | 143 | 144 | 119 | 143 | 144 | 119 | 2002 | 143 | 144 | 119 |
| Donald D. Chamberlin | 44 | 87 | 69 | 44 | 87 | 69 | 2003 | 44 | 87 | 69 |
| Ronald Fagin | 11 | 39 | 32 | 11 | 39 | 32 | 2004 | 11 | 39 | 32 |
| Lowest Ranking Point | 145 | 196 | 183 | 143 | 218 | 222 | | 145 | 196 | 183 |
| Sum of Rank points | 440 | 560 | 498 | 451 | 543 | 539 | | 464 | 494 | 469 |

Table 5: Citations indices of scientists awarded with SIGMOD E. F. Codd Innovations Award.

Table 5 presents the list of the awarded scientists by the '*SIGMOD E.F.Codd Innovations Award*'. In the first group of columns entitled as ($h$, $h_c$, $h_t$) the current position of each scientist is presented by using the respective index. The second triplet of ($h$, $h_c$, $h_t$) shows the scientist positions one year before the awarding. Column *year* shows the year of the awarding, whereas the last triplet of ($h$, $h_c$, $h_t$) shows his position during (at the end of) the year of the awarding.

Although our data are not complete for the time period after 2000, however, we can make interesting observations for the years before 2000:

- **C. Mohan.** Although at this moment he is ranked relatively low by using the *trend h-index* and *contemporary h-index*, during the year of 1996 he was ranked higher according to the *trend h-index*. This was later depicted on the *h-index* since from the $49^{th}$ position where he was ranked during 1996 he now climbed up to the $28^{th}$ position.

- Other similar qualitative cases with obvious difference in the actual ranking are of **Hector Garcia-Molina** and **Philip A. Bernstein**.

- **Serge Abiteboul.** During the year of the awarding the *trend h-index* is relatively low (compared to the *contemporary h-index*). According to the *contemporary h-index*, he was ranked in a higher place. This shows that he had presented interesting work during the age of the awarding; he received the award before his work gets reflected to the *trend h-index* and *h-index*. Thus, in some cases, the *contemporary h-index* gives information that it cannot be depicted into the other indices.

- For the cases of **Michael Stonebraker** and **David J. DeWitt**, we see that they are stable at the top.

## 4.2 Experiments with conferences and journals ranking

### 4.2.1 Experiments with conferences ranking

To evaluate our citation indices on conference ranking, we extract only the database conferences (as defined in [13]) from the data we used in the previous section. In the first part of this section we will make experiments using only the indicators that we fixed for scientists, namely *h-index*, *normalized h-index*, *contemporary h-index* and *trend h-index*. In Table 6 we present the top-10 conferences using the *h-index* for the ordering[6]. Since the quality of the conferences is relatively constant, we observe that in Tables 7 and 8 there are no significant differences in the ranking. The ordering changes dramatically in Table 9 due to the fact that complete data exist only for some conferences.

| Name | $h$ | $a$ | $N_{c,tot}$ | $N_p$ |
|---|---|---|---|---|
| 1.sigmod | 45 | 6.05 | 12261 | 2059 |
| 2.vldb | 37 | 7.10 | 9729 | 2192 |
| 3.pods | 26 | 5.74 | 3883 | 776 |
| 4.icde | 22 | 6.83 | 3307 | 1970 |
| 5.er | 16 | 5.80 | 1486 | 1338 |
| 6.edbt | 13 | 3.89 | 658 | 434 |
| 7.eds | 12 | 3.65 | 527 | 101 |
| 8.adbt | 12 | 2.86 | 412 | 42 |
| 9.icdt | 11 | 4.79 | 580 | 313 |
| 10.oodbs | 11 | 3.96 | 480 | 122 |

Table 6: Conferences ranking using the *h-index*.

| Name | $h_c$ | $a_c$ | $h$ | $N_{c,tot}$ | $N_p$ |
|---|---|---|---|---|---|
| 1.sigmod | 21 | 9.49 | 45 | 12261 | 2059 |
| 2.vldb | 17 | 11.34 | 37 | 9729 | 2192 |
| 3.pods | 12 | 9.73 | 26 | 3883 | 776 |
| 4.icde | 11 | 11.88 | 22 | 3307 | 1970 |
| 5.icdt | 8 | 5.04 | 11 | 580 | 313 |
| 6.edbt | 7 | 6.16 | 13 | 658 | 434 |
| 7.oodbs | 6 | 3.63 | 11 | 480 | 122 |
| 8.er | 5 | 16.21 | 16 | 1486 | 1338 |
| 9.kdd | 5 | 6.89 | 6 | 243 | 1074 |
| 10.dood | 5 | 6.57 | 8 | 440 | 171 |

Table 7: Conferences ranking using the *contemporary h-index*.

| Name | $h_t$ | $a_t$ | $h$ | $N_{c,tot}$ | $N_p$ |
|---|---|---|---|---|---|
| 1.sigmod | 34 | 6.67 | 45 | 12261 | 2059 |
| 2.vldb | 27 | 8.00 | 37 | 9729 | 2192 |
| 3.pods | 19 | 6.53 | 26 | 3883 | 776 |
| 4.icde | 16 | 9.52 | 22 | 3307 | 1970 |
| 5.icdt | 12 | 3.67 | 11 | 580 | 313 |
| 6.edbt | 9 | 6.02 | 13 | 658 | 434 |
| 7.er | 8 | 10.35 | 16 | 1486 | 1338 |
| 8.dood | 8 | 4.43 | 8 | 440 | 171 |
| 9.kdd | 7 | 6.42 | 6 | 243 | 1074 |
| 10.dbpl | 7 | 5.11 | 8 | 410 | 228 |

Table 8: Conferences ranking using the *trend h-index*.

| Name | $h_n$ | $h$ | $a$ | $N_{c,tot}$ | $N_p$ |
|---|---|---|---|---|---|
| 1.adbt | 0.28 | 12 | 2.86 | 412 | 42 |
| 2.dpds | 0.17 | 7 | 2.97 | 146 | 39 |
| 3.eds | 0.11 | 12 | 3.65 | 527 | 101 |
| 4.icod | 0.11 | 6 | 3 | 108 | 52 |
| 5.jcdkb | 0.11 | 8 | 3.32 | 213 | 70 |
| 6.ddb | 0.09 | 4 | 6.87 | 110 | 44 |
| 7.oodbs | 0.09 | 11 | 3.96 | 480 | 122 |
| 8.tdb | 0.08 | 3 | 6.44 | 58 | 36 |
| 9.berkeley | 0.07 | 10 | 3.52 | 352 | 142 |

Table 9: Conferences ranking using the *normalized h-index*.

In Figure 2 we present in the same way we used for scientists, the progress of selected conferences. Note here that the *h-index* is shown per year in the graphs, which means that this is the computed *h-index* during the specific year. E.g., the *h-index* that is computed for the VLDB for 1995 is the *h-index* that is computed if we exclude everything from our database after 1995. Apparently, this is different from a score for the VLDB'95, which we defined earlier as $h_{1995}$.[7]

---

[6]The symbol $a_c$ in Table 6 and the symbol $a_t$ in Table 8 correspond to the *a-index* on Definition 2.

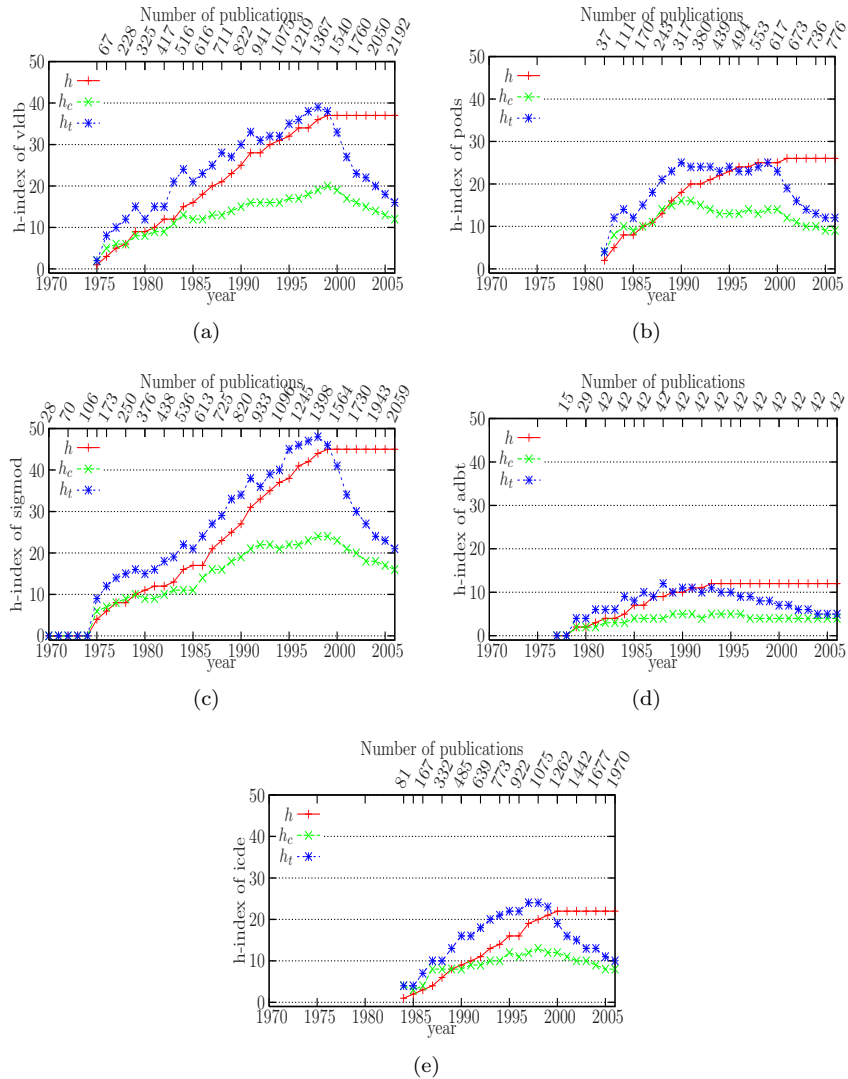

Figure 2: The *h-index* of database conferences.

Figure 2(c) presents the history of the SIGMOD conference. According to the tables, SIGMOD is ranked first. In the figure, we observe its steeply ascending line. Also the *trend h-index* remains higher than the *h-index* (until 1999). On the other hand, the PODS conference (Figure 2(b)) follows a bending line after 1993. Thus, the *h-index* is lightly increased. This can be attributed to the shift in the conference topics; it moved from issues related to the common ground of artificial intelligence and databases, to topics closer to database theory, thus losing some popularity. ICDE is a relatively younger conference compared to the rest of the conferences presented, but we can see in the plot (Figure 2(e)), that it follows a rapidly ascending course, indicating a very competitive conference.

Finally, with respect to the ADBT conference (Figure 2(d)) we mention that this conference was organized only three times (1977, 1979 and 1982). As we can see in the upper $x$ axis, the number of publications stops increasing after 1982. Thus, we can not compare it to the rest of the conferences. The small number of publications of ADBT is the reason that ADBT is ranked first in Table 9.

The next step in conference ranking is to evaluate the usefulness and benefit of Definitions 6 and 7.

---

[7]Due to the lack of citations for the years after 1999, in all graphs there is a stabilization of the *h-index* line and a downfall for the indicators *trend h-index* and *contemporary h-index*.

This way, we evaluate, for example, VLDB'95 independently from VLDB'94. Obviously, in this case it is meaningless to add a second time dimension (with indicators *contemporary h-index* and *trend h-index*). The *contemporary h-index* of VLDB'95 will be stable during all the following years, since all papers are published during the same year. On the other hand, it is not important to see whether a conference organized in 1980 still gets citations.

Indicatively, we present the Tables 10 and 11 which present the conferences ranking for the years 1995 and 1990 respectively. In part (a) of these tables the ordering is performed by using the *yearly h-index* ($h_y$). Factor $a$ is the second criterion for the ranking. We also present the columns $h_y^n$, which is the *h-index* divided by the number of publications $N_{p,y}$. Also, column $N_{c,1995}$ is the number of citations to papers published during 1995. In the second part (b) of the tables, the ordering is computed based to the *normalized h-index*. Notice here, that although it seems to have equivalences by using $h_y^n$, the real numbers make such a situation almost unprovable (i.e., $5/24 = 0.20833$, $6/29 = 0.206897$).

What we observe here, is that there are no important differences in the ranking for the two indicative years, neither using the *normalized h-index*. On the other hand, it is obvious that the use of *normalized h-index* gives a small advantage to the conferences that have a small number of publications. For example, the VLDB conference is almost stable in the first place using the *yearly h-index*, but it is improbable to get the first place using the *normalized yearly h-index*.

| Name | $h_{1995}$ | $a$ | $h_{1995}^n$ | $N_{c,1995}$ | $N_{p,1995}$ |
|---|---|---|---|---|---|
| 1.vldb | 11 | 3.57 | 0.15 | 432 | 72 |
| 2.sigmod | 9 | 4.62 | 0.10 | 375 | 85 |
| 3.icde | 6 | 6.63 | 0.08 | 239 | 68 |
| 4.pods | 6 | 4.16 | 0.20 | 150 | 29 |
| 5.ssd | 5 | 2.08 | 0.20 | 52 | 24 |
| 6.kdd | 4 | 3.81 | 0.07 | 61 | 56 |
| 7.cikm | 3 | 6.22 | 0.05 | 56 | 55 |
| 8.dood | 3 | 5.88 | 0.06 | 53 | 46 |
| 9.icdt | 3 | 3.66 | 0.08 | 33 | 34 |
| 10.er | 3 | 3.33 | 0.06 | 30 | 47 |

(a)

| Name | $h_{1995}^n$ | $h_{1995}$ | $N_{p,1995}$ |
|---|---|---|---|
| 1.ssd | 0.20 | 5 | 24 |
| 2.pods | 0.20 | 6 | 29 |
| 3.cdb | 0.2 | 2 | 10 |
| 4.vldb | 0.15 | 11 | 72 |
| 5.coopis | 0.14 | 3 | 21 |
| 6.artdb | 0.11 | 2 | 17 |
| 7.sdb | 0.11 | 1 | 9 |
| 8.sigmod | 0.10 | 9 | 85 |
| 9.ride | 0.10 | 2 | 19 |
| 10.tdb | 0.1 | 2 | 20 |

(b)

Table 10: Conferences ranking for the year 1995.

| Name | $h_{1990}$ | $a$ | $h_{1990}^n$ | $N_{c,1990}$ | $N_{p,1990}$ |
|---|---|---|---|---|---|
| 1.vldb | 16 | 2.57 | 0.26 | 659 | 60 |
| 2.sigmod | 15 | 3.44 | 0.31 | 776 | 48 |
| 3.icde | 11 | 2.76 | 0.16 | 335 | 67 |
| 4.pods | 11 | 2.40 | 0.30 | 291 | 36 |
| 5.edbt | 7 | 2.83 | 0.21 | 139 | 32 |
| 6.icdt | 5 | 4.32 | 0.14 | 108 | 34 |
| 7.dpds | 4 | 3.75 | 0.22 | 60 | 18 |
| 8.er | 3 | 4.66 | 0.08 | 42 | 35 |
| 9.ds | 3 | 4.11 | 0.12 | 37 | 24 |
| 10.ssdbm | 3 | 3 | 0.16 | 27 | 18 |

(a)

| Name | $h_{1995}^n$ | $h_{1995}$ | $N_{p,1995}$ |
|---|---|---|---|
| 1.sigmod | 0.31 | 15 | 48 |
| 2.pods | 0.30 | 11 | 36 |
| 3.vldb | 0.26 | 16 | 60 |
| 4.dpds | 0.22 | 4 | 18 |
| 5.edbt | 0.21 | 7 | 32 |
| 6.ssdbm | 0.16 | 3 | 18 |
| 7.icde | 0.16 | 11 | 67 |
| 8.icdt | 0.14 | 5 | 34 |
| 9.ds | 0.12 | 3 | 24 |
| 10.ewdw | 0.10 | 3 | 29 |

(b)

Table 11: Conferences ranking for the year 1990.

In Figure 3 we present the plots for the values of *yearly h-index* ($h_y$) and *normalized yearly h-index* ($h_y^n$) for the top four conferences VLDB, PODS, SIGMOD and ICDE. The values for $h_y$ are drawn using bars, because each value is independent from the rest ones. The value for $h_y$ of a conference has different upper bound for each year. The upper bound for each year is defined by the number of papers published during this year. This is depicted on the upper $x$ axes. On the other hand, the $h_y^n$ values are normalized. So, it is a comparable value for the two years of a conference and it is drawn with the (red) cross points line. The values for the $h_y^n$ index are presented in axes $y2$. There is no association of axes $y1$ to $y2$, thus we cannot compare (obviously) the values of $h_y^n$ to $h_y$. The only remark that we can make is that the one curve follows approximately the other. This comes in agreement with the conclusions derived from Tables 10 and 11.

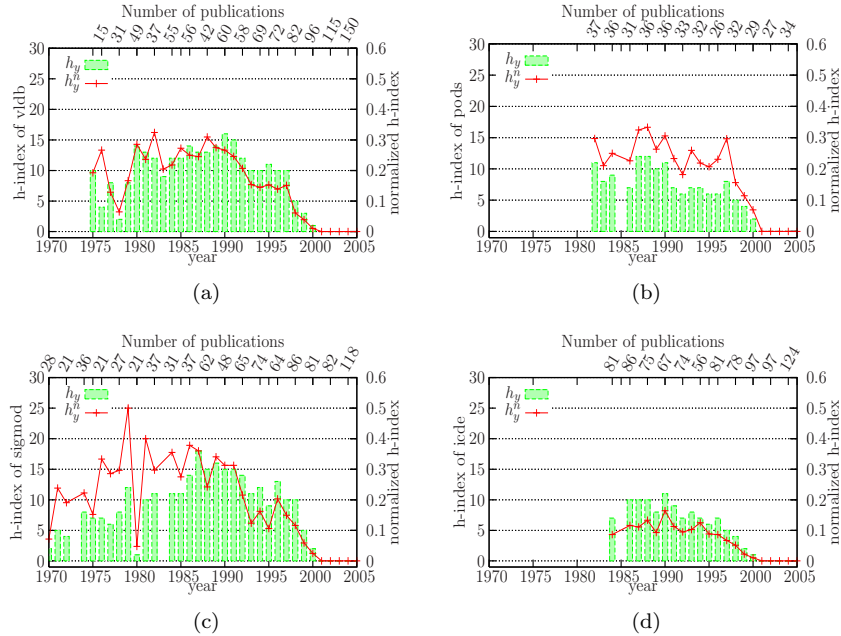

Figure 3: The *yearly h-index* and *normalized yearly h-index* of database conferences.

### 4.2.2 Experiments with journals ranking

In the case of journals, we can use the basic form of *h-index* as well as the generalizations *contemporary h-index* and *trend h-index* and the variant *normalized h-index* we defined for scientists and for conferences. Here, similarly to the case of conferences, the *normalized h-index* is a valuable indicator contrary to the case of the scientists.

Tabular data in Figures 4(a), 4(b), 5(a) and 5(b) present the top-10 journals according to the four aforementioned indices. As expected, the ACM TODS (tods), IEEE TDKE (tkde), SIGMOD Record (sigmod) are the top three journals. The striking observation is that the Information Systems (is) drops in the ranking with the *contemporary h-index* and *trend h-index* as compared to its position with *h-index*, implying that it is not considered an exceptionally prestigious journal anymore. On the contrary, SIGMOD Record and VLDB Journal (vldb) show an uprising trend.

In Figure 6 we present the results of computing the defined indices for the major journals of the database domain on a per year basis. Due to the lack of available data after the year 2000, all indices drop steeply. Though, the case of ACM TODS is worthwhile mentioning. Its *trend h-index* drops after 1993, which can be attributed to the relatively large end-to-end publication time of its articles during the years 1990-2000 [33], which acted as an impediment for the authors to submit their works in that venue. Fortunately, this is not the case anymore. Also, the case of SIGMOD Record is characteristic, because, even though it is published since 1970, its indices get really noticeable only after 1980, when this newsletter started to publish some very good survey-type articles and was freely available on the Web, which improved its visibility.

Finally, in Figure 7 we present the results of computing the *yearly h-index* and the *normalized yearly h-index* for the major journals of the database domain on a per year basis. In these figures we can easily see that the ACM TODS journal undoubtedly gets the first place. Also, Figure 7(a) shows that the *yearly h-index* follows a decreasing path which comes in agreement with Figure 6(a). Also, Figure 7(e) comes in aggrement with Figure 6(e). In Figure 7(c) it is shown that the small number of published papers for the years 1984 and 1986 make the *normalized yearly h-index* to get higher than usual for the SIGMOD Record journal. TKDE (Figure 7(b)) seems to follow a descreasing slope and finally, VLDB (Figure 7(d)) follows an uprising trend until 1996.

| Name | $h$ | $a$ | $N_{c,tot}$ | $N_p$ |
|---|---|---|---|---|
| 1.tods | 49 | 3.88 | 9329 | 598 |
| 2.tkde | 18 | 4.69 | 1520 | 1388 |
| 3.is | 16 | 4.71 | 1208 | 934 |
| 4.sigmod | 15 | 5.07 | 1142 | 1349 |
| 5.tois | 13 | 4.37 | 740 | 378 |
| 6.debu | 11 | 7.13 | 863 | 877 |
| 7.vldb | 9 | 5.03 | 408 | 281 |
| 8.ipl | 8 | 6.06 | 388 | 4939 |
| 9.dke | 6 | 8.77 | 316 | 773 |
| 10.dpd | 6 | 5.25 | 189 | 238 |

(a) Journal ranking using the *h-index*.

| Name | $h_n$ | $h$ | $a$ | $N_{c,tot}$ | $N_p$ |
|---|---|---|---|---|---|
| 1.tods | 0.08 | 49 | 3.88 | 9329 | 598 |
| 2.tois | 0.03 | 13 | 4.37 | 740 | 378 |
| 3.vldb | 0.03 | 9 | 5.03 | 408 | 281 |
| 4.dpd | 0.02 | 6 | 5.25 | 189 | 238 |
| 5.jiis | 0.01 | 6 | 4.33 | 156 | 318 |
| 6.datamine | 0.01 | 3 | 5.11 | 46 | 162 |
| 7.is | 0.01 | 16 | 4.71 | 1208 | 934 |
| 8.ijcis | 0.01 | 4 | 3.12 | 50 | 255 |
| 9.tkde | 0.01 | 18 | 4.69 | 1520 | 1388 |
| 10.debu | 0.01 | 11 | 7.13 | 863 | 877 |

(b) Journal ranking using the *normalized h-index*.

Figure 4: Journal ranking using *h-index* (left) and *normalized h-index* (right).

| Name | $h_c$ | $a_c$ | $h$ | $N_{c,tot}$ | $N_p$ |
|---|---|---|---|---|---|
| 1.tods | 18 | 6.25 | 49 | 9329 | 598 |
| 2.tkde | 10 | 6.40 | 18 | 1520 | 1388 |
| 3.sigmod | 9 | 6.17 | 15 | 1142 | 1349 |
| 4.debu | 6 | 9.21 | 11 | 863 | 877 |
| 5.vldb | 6 | 6.47 | 9 | 408 | 281 |
| 6.tois | 6 | 6.09 | 13 | 740 | 378 |
| 7.is | 5 | 12.77 | 16 | 1208 | 934 |
| 8.dpd | 5 | 4.19 | 6 | 189 | 238 |
| 9.jiis | 5 | 3.79 | 6 | 156 | 318 |
| 10.dke | 4 | 7.70 | 6 | 316 | 773 |

(a) Journal ranking using the *contemporary h-index*.

| Name | $h_t$ | $a_t$ | $h$ | $N_{c,tot}$ | $N_p$ |
|---|---|---|---|---|---|
| 1.tods | 28 | 4.93 | 49 | 9329 | 598 |
| 2.tkde | 13 | 6.64 | 18 | 1520 | 1388 |
| 3.sigmod | 12 | 5.85 | 15 | 1142 | 1349 |
| 4.vldb | 10 | 3.75 | 9 | 408 | 281 |
| 5.is | 9 | 7.11 | 16 | 1208 | 934 |
| 6.debu | 9 | 6.98 | 11 | 863 | 877 |
| 7.tois | 9 | 4.83 | 13 | 740 | 378 |
| 8.dpd | 6 | 4.88 | 6 | 189 | 238 |
| 9.jiis | 6 | 4.75 | 6 | 156 | 318 |
| 10.dke | 5 | 8.18 | 6 | 316 | 773 |

(b) Journal ranking using the *trend h-index*.

Figure 5: Journal ranking using *contemporary h-index* (left) and *trend h-index* (right).

# 5 Conclusions

Estimating the significance of a scientist's work is a very important issue for prize awarding, faculty recruiting, etc. This issue has received a lot of attention during the last years and a number of metrics have been proposed which are based on arithmetics upon the number of articles published by a scientist and the total number of citations to these articles. The interest on these topics has been renewed and in a path-breaking paper, J. E. Hirsch proposed the *h-index* to perform fair ranking of scientists, avoiding many of the drawbacks of the earlier bibliographic ranking methods.

The initial proposal and meaning of the *h-index* has various shortcomings, mainly of its inability to differentiate between active and inactive (or retired) scientists and its weakness to differentiate between significant works in the past (but not any more) and the works which are "trendy" or the works which continue to shape the scientific thinking.

Based on the identification of these shortcomings of *h-index*, we proposed in this article a number of effective *h-index* generalizations and some variants. Some of these novel citation indices aim at the ranking of scientists by taking into account the age of the published articles, or the age of the citations to each article. The other citations indices aim at ranking publication venues, i.e., conferences and journals, taking into account the variable number of published articles.

To evaluate the proposed ranking metrics, we conducted extensive experiments on an online bibliographic database containing data from journal and conference publications as well, and moreover focused in the specific area of databases. From the results we obtained, we concluded that *h-index* is not a general purpose indicative metric. Some of the novel indices, namely *contemporary h-index* and *trend h-index*, are able to disclose latent facts in citation networks, like trendsetters and brilliant young scientists. For the case of conference and journal ranking, the indices *normalized h-index*, *contemporary h-index* and *trend h-index* give a more fair view for the ranking. Finally, the *yearly h-index* and the *normalized yearly h-index* can be used in order to evaluate separately each conference/journal's success.

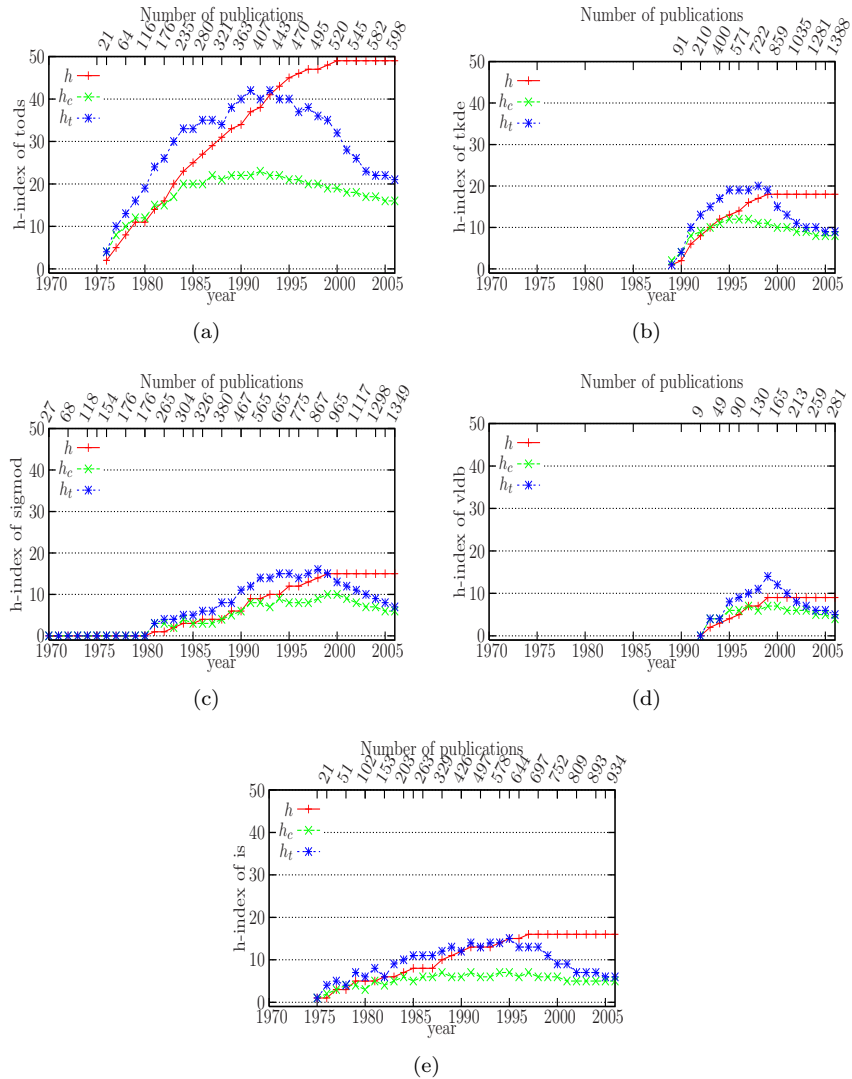

Figure 6: The *h-index*, *contemporary h-index* and *trend h-index* of major database journals.

# References

[1] P. Ball. Index aims for fair ranking of scientists – *h*-index sums up publication record. *Nature*, 436(7053):900, 2005.

[2] J. Bar-Ilan. *h*-index for price medalists revisited. *ISSI Newsletter*, 5, Jan. 2006.

[3] S. J. Barnes. Assessing the value of IS journals. *Communications of the ACM*, 48(1):110–112, 2005.

[4] P. D. Batista, M. G. Campiteli, O. Kinouchi, and A. S. Martinez. Is it possible to compare researchers with different scientific interests? *Scientometrics*, 68(1), 2007. To appear.

[5] P. A. Bernstein, E. Bertino, A. Heuer, C. J. Jensen, H. Meyer, M. Tamer Ozsu, R. T. Snodgrass, and K.-Y. Whang. An apples-to-apples comparison of two database journals. *ACM SIGMOD Record*, 34(4):61–64, 2005.

[6] P. Bharati and P. Tarasewich. Global perceptions of journals publishing e-commerce research. *Communications of the ACM*, 45(5):21–26, 2002.

[7] J. Bollen, M. A. Rodriguez, and H. Van de Sompel. Journal status, May 2006. Preprint available at http://arxiv.org/abs/cs/0601030.

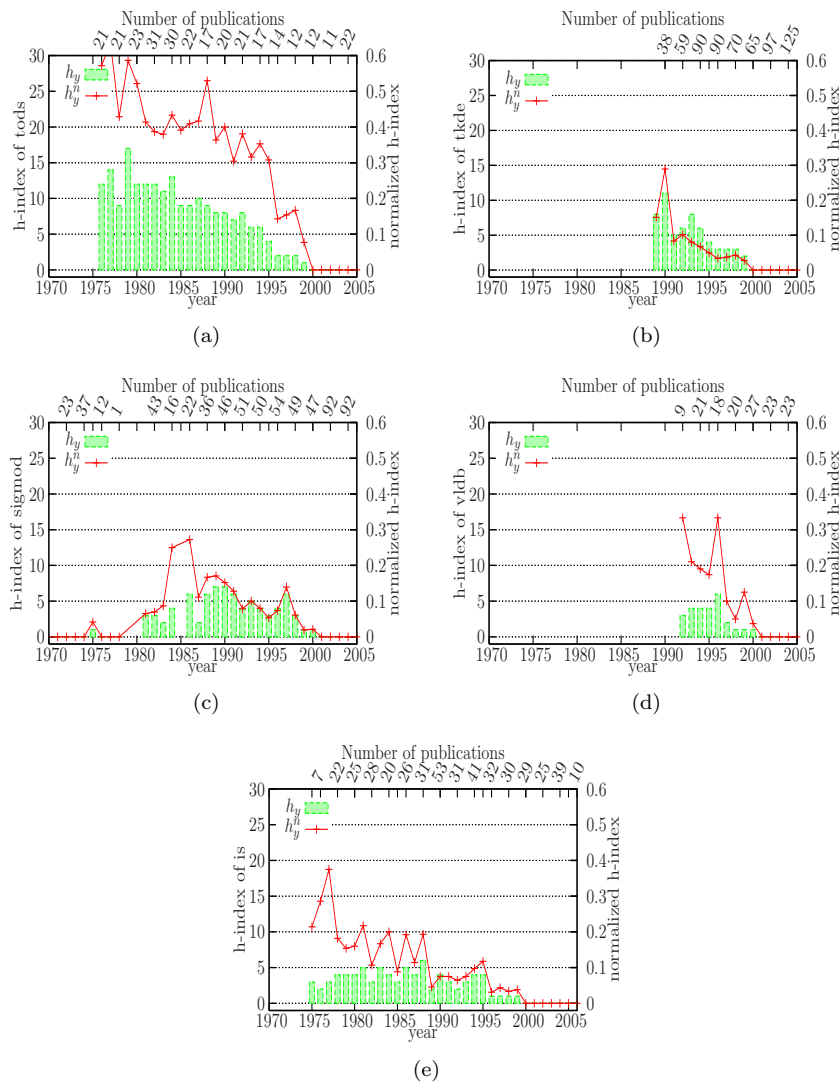

Figure 7: The *yearly h-index* and *normalized yearly h-index* of major database journals.

[8] L. Bornmann and H.-D. Daniel. Does the *h*-index for ranking of scientists really work? *Scientometrics*, 65(3):391–392, 2005.

[9] T. Braun, W. Glanzel, and A. Schubert. A Hirsch-type index for journals. *The Scientist*, 19(22):8–10, 2005.

[10] B. Cronin and L. Meho. Using the H-index to rank influential information scsientists. *Journal of the American Society for Information Science and Technology*, 57(9):1275–1278, 2006.

[11] L. Egghe. Dynamic *h*-index: The Hirsch index in function of time. *Scientometrics*, 2006. To appear.

[12] L. Egghe. Theory and practise of the *g*-index. *Scientometrics*, 2006. To appear.

[13] E. Elmacioglu and D. Lee. On six degrees of separation in DBLP-DB and more. *ACM SIGMOD Record*, 34(2):33–40, 2005.

[14] E. Garfield. Citation analysis as a tool in journal evaluation. *Science*, 178:471–479, 1972.

[15] B. C. Hardgrave and K. A. Walstrom. Forums for MIS scholars. *Communications of the ACM*, 40(11):119–124, 1997.

[16] J. E. Hirsch. An index to quantify an individual's scientific research output. *Proceedings of the National Academy of Sciences*, 102(46):16569–16572, 2005.

[17] C. W. Holsapple, H. Johnson, L. E.and Manakyan, and J. Tanner. Business computing research journals: A normalized citation analysis. *Journal of Management Information Systems*, 11(1):131–140, 1994.

[18] P. Katerattanakul, B. T. Han, and S. Hong. Objective quality ranking of computing journals. *Communications of the ACM*, 46(10):111–114, 2003.

[19] R. Kelly Rainer and M. D. Miller. Examining differences across journal rankings. *Communications of the ACM*, 48(2):91–94, 2005.

[20] J. Kleinberg. Authoritative sources in a hyperlinked environment. *Journal of the ACM*, 46(5):604–632, 1999.

[21] P. Lowry, D. Romans, and A. Curtis. Global journal prestige and supporting disciplines: A scientometric study of information systems journals. *Journal of the Association for Information Systems*, 5(2):29–75, 2004.

[22] N. A. Mylonopoulos and V. Theoharakis. Global perception of IS journals. *Communications of the ACM*, 44(9):29–33, 2001.

[23] M.A. Nascimento, J. Sander, and J. Pound. Analysis of SIGMOD's co-authorship graph. *ACM SIGMOD Record*, 32(3):8–10, 2003.

[24] S. P. Nerur, R. Sikora, G. Mangalaraj, and V. Balijepally. Assessing the relative influence of journals in a citation network. *Communications of the ACM*, 48(11):71–74, 2005.

[25] L. Page, S. Brin, R. Motwani, and T. Winograd. The PageRank citation ranking: Bringing order to the Web. Technical Report TR-1999-66, Stanford University, 1999.

[26] A. F. J. van Raan. Comparison of the Hirsch-index with standard bibliometric indicators and with peer judgment for 147 chemistry research groups. *Scientometrics*, 67(3):491–502, 2006.

[27] E. Rahm and A. Thor. Citation analysis of database publications. *ACM SIGMOD Record*, 34(4):48–53, 2005.

[28] R. Rousseau. A case study: Evolution of JASIS' Hirsch index. *Library and Information Science*, Jan. 2006. Available at http://eprints.rcils.org/archive/00005430.

[29] R. B. Schwartz and M. C. Russo. How to quickly find articles in the top IS journals. *Communications of the ACM*, 47(2):98–101, 2004.

[30] A. Sidiropoulos and Y. Manolopoulos. A citation-based system to assist prize awarding. *ACM SIGMOD Record*, 34(4):54–60, 2005.

[31] A. Sidiropoulos and Y. Manolopoulos. A new perspective to automatically rank scientific conferences using digital libraries. *Information Processing & Management*, 41(2):289–312, 2005.

[32] A. Sidiropoulos and Y. Manolopoulos. Generalized comparison of graph-based ranking algorithms for publications and authors. *Journal for Systems and Software*, 2006. To appear.

[33] R. Snodgrass. Journal relevance. *ACM SIGMOD Record*, 32(3):11–15, 2003.